\documentclass[preprint,12pt]{elsarticle}

\usepackage[utf8]{inputenc}
\usepackage{amsmath}
\usepackage{amsfonts}
\usepackage{amssymb}
\usepackage{empheq}
\usepackage{graphicx}
\usepackage[left=2cm,right=2cm,top=2cm,bottom=2cm]{geometry}

\usepackage{hyperref}
\usepackage[table,xcdraw]{xcolor}

\journal{Optik}

\begin{document}

\begin{frontmatter}

\title{{\bf Optimum design of permeable diffractive lenses \\ based on photon sieves}}

\author[1]{Veronica Pastor-Villarrubia}
\author[2]{Angela Soria-Garcia}
\author[2]{Joaquin Andres-Porras}
\author[2]{Jesus del Hoyo}
\author[1,3]{Mahmoud H. Elshorbagy}
\author[2]{Luis Miguel Sanchez-Brea}
\author[1]{Javier Alda}

\address[1]{Applied Optics Complutense Group, Optics Departament, Faculty of Optics and Optometry, Universidad Complutense de Madrid. C/ Arcos de Jalón, 118. 28037 Madrid. Spain.}

\address[2]{Applied Optics Complutense Group, Optics Departament, Faculty of Physics, Universidad Complutense de Madrid. Ciudad Universitaria s/n. 28040 Madrid. Spain}

\address[3]{Physics Department, Faculty of Science, Minia University. El Minia. 61519 Egypt.}

\address{Corresponding author: Veronica Pastor-Villarrubia. E-mail: veronica.pastor@ucm.es}

\begin{abstract}
Photon sieves are permeable  diffractive optical elements  generated by open apertures on a substrate. 
These elements are well suited for the monitoring of running fluids. 
Our analysis considers the fabrication constrains of the photon sieve and translate them into values of the optical parameters of the element.
When used as focusing elements,  or diffractive lenses, the spatial distribution of apertures can be designed to maximize the intensity at the focal plane and the permeability of the device. This is done by defining a weighted merit function. 
The computation time  of this merit function is key when applying different strategies for the design, which often require a very large number of calculations of this merit function. Then, besides using a reliable 
propagation method, we have included an analytic solution applicable for circular apertures. Also, a geometrical 
merit function is proposed to simplify and reduce the computation even more. 
The methods proposed in this contribution are compared in terms of the focused irradiance and permeability parameters, allowing an educated choice adapted to the given case or application.
In this contribution we analyze several methods to generate photon sieves in an optimum manner.
The resulted spatial distributions resemble the classical Fresnel zone arrangement.  
\end{abstract}

\begin{keyword}
Diffraction \sep Diffractive Optical Elements \sep Optimization  \sep Permeable Optics \sep Photon Sieve
\end{keyword}

\end{frontmatter}

\section{Introduction}

Since long ago, diffractive optics have been routinely used to expand and improve the capabilities of optical systems. Among them, Fresnel Zone Plates (FZP) mimic the focusing properties of refractive and reflective optics, generating lightweight and compact designs \cite{Moreno1997,Cao2004, Davis2007,Reid2009}.
There are many types of Fresnel lenses depending on how the transmittance is modulated: amplitude, phase, or a combination of both \cite{Snigirev1996,VilaComamala2006, Li2018}. Moreover, polarization elements can be selectively added to the Fresnel arrangement to obtain new capabilities as true vectorial diffractive elements \cite{SoriaGarcia2023,TorcalMilla2022}.
One important advantage of Frensel lens is its flat shape that allows to directly write them on a plane plate. 
In any case, as it happens with refractive optics, Diffractive Optical Elements (DOEs) are solid and non permeable objects that physically separate two spatial regions, precluding the interchange of material through them. This is where a Photon Sieve (PS) helps to design permeable elements that could be of application as an optical system immersed within a running fluid (liquid or gas). Therefore, a PS can be seen as a collection of clear apertures that are arranged to perform as a DOE. 

So far, PS have been applied to the design of lenses in the X-ray range \cite{Snigirev1996, kipp_xrayps_nature2001}, as large aperture spatial instrumentation \cite{andersen_pstelescope_spie2006,Gao2008}, or as advanced optical elements 
\cite{sabatyan_axilensPS_appopt2014,sabatyan_apodizedPS_appopt2012}. 
In this contribution, our interest is focused on its use as optical elements in immersed optics for the monitoring of running fluids \cite{PastorVillarrubia2023,PastorVillarrubiaOPTOEL}. One of the options for this application uses  a photon sieve operating in reflection, which may be functionalized in the illuminated surface to respond to the presence of a given specimen or material \cite{ArroyoHernandez2003,Gopinath2012,Kaushik2020}. 

The key point of a successful design relies on the adequate arrangement of apertures on the surface of the photon sieve \cite{sabatyan_gaussiandPS_appopt2011}. 
  However, depending on the available fabrication tools, the apertures can take arbitrary shapes while maintaining the PS physical integrity.  
  Our approach has been limited to the case of circular apertures. The reason for this choice is the easiness of its fabrication by micro-mechanical or laser-assisted drilling \cite{wang_laserdrilling_appphya2010,geng_laserdrilling_ijamt2016}. 
  Besides, when considering running fluids, circular aperture disturb less the currents than those apertures showing sharp geometries \cite{tanner_flowapertures_jnmhff2019}.
Our selection has also served to better understand how the fabrication constrains affect to the optical characteristics of the element, in terms of its usable aperture, $F_{\#}$, and focal length.  To properly show the advantages and limitations of our approach,  section \ref{sec:fundamentals} presents the basic concepts used in the design of Fresnel lens realized as a photon sieve, including those conditions derived from the fabrication constrains. After this discussion, section \ref{sec:optimization}  explains several definitions of the merit function used in the generation of optimum photon sieves.
 In our case, we consider both the optical performance, and the permeability of the element. 
 Then, in section \ref{sec:generation}, we present four methods to design focusing photon sieves. Some of them are deterministic, providing the same design once the merit function is set, and some others change from one realization to the next, producing  variations in the aperture's arrangement. 
Section \ref{sec:analysis} makes a comparative analysis of these methods in term of the obtainable values of 
the parameters of interest, and the computational speed.
Finally, the main conclusions of this contribution are presented in section \ref{sec:conclusions}.

\section{Fundamentals and limitations of photon sieves}
\label{sec:fundamentals}

A photon sieve is a special type of DOE where the mask contains clear apertures
which allow light and matter pass through them. 
Therefore, they are truly permeable elements that focus light using diffraction. 
Our photon sieves should behave as lenses, where the main optical parameter are the focal length, $f'$,  and  diameter. The FZP corresponds to the basic geometry of a diffractive lens, where 
the constructive contribution between Fresnel zones generates the desired focusing capabilities of the element. 
It is well known \cite{Hecht,DiffOptFZP} that, for a given focal length, $f'$, and wavelength, $\lambda$, the  radii of the Fresnel zones are given as:
\begin{equation}
    R_{{\rm Fresnel}, m} = \sqrt{m \lambda f^\prime + \frac{m^2 \lambda^2}{4} } \approx \sqrt{m \lambda f^\prime}
    \label{eq:FZradius}
,\end{equation}
where $m$ is the order of the zone  that runs from 1 to $M$, being $M$ the order of the last Fresnel zone. 
In Eq. (\ref{eq:FZradius}) we have assumed that $M \ll 4f'/\lambda$ to safely use the approximate relation. At this point we should explain how the index of refraction of the fluid where the photon sieve is immersed in affects the focusing of the device. The wavelength included in Eq. (\ref{eq:FZradius}) can be written in terms of the wavelength in vacuum as $\lambda = \lambda_0/n$, where $n$ is the index of refraction of the fluid. Using this relation, the proposed design can be adapted to the given case.

Once the focal length is set, we wonder how large can be the aperture. As we will see, this limitation is 
related with fabrication constraints  of the photon sieve. 
From Eq. (\ref{eq:FZradius}), the radius of the last zone $M$ defines
the size of the usable aperture, $R_{{\rm Fresnel,}M}=\sqrt{M\lambda f'}$, and therefore, the f-number,  $F_{\#} = f'/D$, is also obtained as
\begin{equation}
F_{\#}= \sqrt{\frac{f'}{4M\lambda}}
\label{eq:Fnumber}
.\end{equation}
Therefore, when considering the fabrication limits, we need to obtain the minimum width of the Fresnel zone in the lens, that corresponds with the last zone, resulting:
\begin{equation}
\Delta R_{{\rm Fresnel,}M} = R_{{\rm Fresnel,}M} - R_{{\rm Fresnel,}M-1} \simeq \sqrt{\frac{\lambda f'}{4M}}
,\label{eg:width_minima}
\end{equation}
that combined with Eq. (\ref{eq:Fnumber}) gives a quite simple relation between $F_{\#}$ and the width of the last zone: $\Delta R_{{\rm Fresnel,}M}= F_{\#} \lambda$. Here is where the fabrication methodology plays an important role.
If we assume that the photon sieve is generated using a collection of circular open apertures, which, in a simple 
distribution should be located selectively on the Fresnel zones, the minimum fabricable radius, $R_{\rm min}$, should fit 
within the narrowest zone, i.e., $2R_{\rm min} = \Delta R_{{\rm Fresnel,}M}$. This condition can be written as
\begin{equation}
R_{\rm min} = \frac{ F_{\#} \lambda}{2}
\label{eq:rho_min}
,\end{equation}
meaning that the f-number of the device  linearly depends on the radius of the minimum fabricable aperture. 
Typically, the size of the mimimun fabricable aperture is larger than the wavelength. Then, the aperture number of fabricable devices will be larger, or much larger, than 1. 
In fact, if the apertures are subwavelength in size, the model and simulation of the system should use the tools of  computational electromagnetism.
Once we have obtained the smallest aperture of the photon sieve, we may define the largest one, $R_{\rm max}$. Our designs start with an open aperture at the center of the PS. The size of this central aperture is given from Eq. (\ref{eq:FZradius}) when $m=1$, as $R_{{\rm Fresnel},1}= \sqrt{\lambda f'}$. Therefore, we may write
\begin{equation}
    R_{\rm max} = R_{{\rm Fresnel},1} =  \sqrt{\lambda f'}
    \label{eq:rho_max}
.\end{equation}
Using Eqs. (\ref{eq:rho_min}) and (\ref{eq:rho_max}) we fixed the limits for the size of the clear apertures. In our analysis, the
only aperture  that is fully determined by the values of the operating wavelength, $\lambda$, and focal length, $f'$, is the first one that is centered at the optical axis of the photon sieve. 
The number, location, and size of the rest of apertures should be part of an optimization process that will be treated in the next section.

Another fabrication constrain is related with the separation between adjacent clear apertures, that we consider as open circles. This limitation is required to
avoid the intersection among apertures and maintain the integrity of the PS mask. 
Therefore, we choose  a minimum distance between holes denoted as $g$. 
This value depends on the fabrication methodology and
the properties of the material of the mask. This means that the distance between the centers of two adjacent holes with radius $R_i$ and $R_j$, should be greater than $R_i+R_j+g$. This condition can be written as:
\begin{equation}
    \sqrt{(x_{0,i}-x_{0,j})^2 + (y_{0,i}-y_{0,j})^2} \geq R_i + R_j + g
,\label{eq:fabricationcondition}
\end{equation}
where $(x_{0,i},y_{0,i})$ and $(x_{0,j},y_{0,j})$ are the coordinates of the hole centers at the plane of the mask, $XY$, with radius $R_i$ and $R_j$, respectively.

Once these constrains have been established, we  may see that  the most relevant aperture for a focusing photon sieve is the central one. In our case, this aperture coincides with the first Fresnel zone, having a radius $R_{{\rm Fresnel,}1} = \sqrt{\lambda f'}$. The rest of circular aperture should contribute to the total irradiance as constructively as possible. In any case, following the results presented in \cite{Goodman2005,Schmerr1998}, the electric field amplitude at a plane parallel to the photon sieve, and located at a distance $z$ from the plane of the mask, is given as:
\begin{equation}
E (x',y') = C \sum_{j=1}^N \pi R_j^2 \exp \left( i \frac{\rho_j^2}{2z} \right) 
\left[  
\frac{2 J_1\left(  k R_j \rho_j  /z  \right)}{ k R_j \rho_j  /z}  \right]
\label{eq:amplitudemultiplecircular}
,\end{equation}
where $C$ is a constant, $k= 2\pi/\lambda$, $R_j$ is the radius of the hole, and $\rho_j=\sqrt{(x'-x_{0,j})^2+(y'-y_{0,j})^2} $, being $(x_{0,j}, y_{0,j})$ the
coordinates of the center of the circular aperture $j$ in the plane $XY$. This electric field is calculated with respect to a coordinate system having axis $X'Y'$, parallel to $XY$ and located at a distance $z$. The sum in Eq. (\ref{eq:amplitudemultiplecircular}) runs for the $N$ circular apertures opened in the photon sieve. 
The irradiance distribution a the focal plane of the PS,  $z=f'$, will be given from Eq. (\ref{eq:amplitudemultiplecircular}) as
\begin{equation}
I(x',y') =  \frac{n}{2 \mu c}\left| E(x',y') \right|^2
\label{eq:irradiancedistribution}
,\end{equation}
where $c$ is the speed of light in vacuum, and  $n$  and $\mu$ are  the index of refraction and the magnetic permeability of the medium where the beam propagates, respectively. The magnetic permeability, $\mu$, should not be mistaken with the permeability of the photon sieve. Besides this analytical approach, the irradiance distribution at the plane of interest, $X'Y'$, can be evaluated using well-founded propagation methods \cite{paper:FastFourierShen}, as those implemented in ``Diffractio" \cite{Diff,Proceeding:Diffractio}. In the cases treated in this paper, we have paid special attention to the fulfillment of the conditions necessary to apply the propagation algorithms.

On the other hand, the permeability of the photon sieve is determined by the open area of the circular holes in the mask. This permeability describes the capability of the system to allow fluids go through the optical system. This total area is given by
\begin{equation}
A_{\rm open} = \sum_{j=1}^N \pi R_j^2
    \label{eq:permeability}
,\end{equation}
A normalized version of the encircled energy around the focal point, and a normalized variation of $A_{\rm open}$  will be determinant to define a merit function for the optimization of the design.

\begin{figure}
\centering
\includegraphics[width=0.95\textwidth]{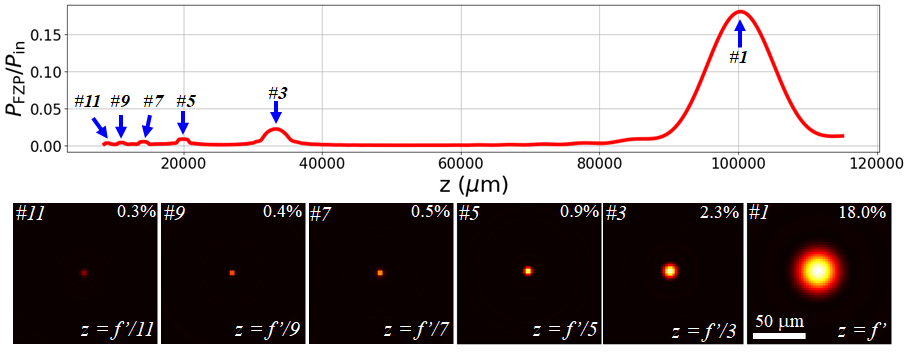}
\caption{
Top: Relative encircled energy with respect to the total energy entering the aperture of the FZP, as a function of the location at the propagation axis, $z$. The arrows and the \#$m$ label represent the location of the higher-order foci, located at a distance $z=f’/m$, where $m$ is an odd number. Bottom: Focal region of the irradiance map at each foci represented in the top plot. The range in the colormap is the same for all the maps and has been normalized to present the full range when $z=f’$, case \#1. The percentage values replicate the relative encircled energy presented in the top figure. }
\label{fig:highorderfoci}

\end{figure}

There exists an important radiometric limitation when considering binary diffractive Fresnel-like lenses, that is related with the occurrence of higher-order focal points. This effect splits the incoming energy among these focal points. As a reference, we have considered the ideal FZP having the same focal distance, $f^{\prime}$, of the designed photon sieves. This diffractive lens shows diffractive focal points at $f^{\prime}/m$, being $m$ and odd number. For this case, we have calculated the ratio between the encircled energy within the Airy circle (as defined in Eq. (\ref{eq:I_airy})) and the incoming energy integrated over the aperture of the lens.
Figure \ref{fig:highorderfoci} shows the evolution of this ratio along the propagation axis,  $z$.
It reveals the occurrence of the higher-order foci. This figure also contains the irradiance maps for the focal planes until $m=11$. We have checked that the first-order focal point, $f^\prime$, is the brightest one, and contains 18.0~\% of the incoming energy for the reference FZP. As the FZP is taken as the reference in the definition of the relative encircle power (see Eq. (\ref{eq:I_normalized})), if we want to obtain the values of the encircled energy with the respect to the incoming energy entering through the aperture of the lens, we should multiply the values of $\hat{P}$ by 0.18.

\section{Optimization}
\label{sec:optimization}

Every optimized design is typically driven by a merit function that is defined to maximize the performance of the design. 
In the case of photon sieves, we define the performance both optically and morphologically.  The morphological portion of the merit function describes its permeability, which is derived from Eq.(\ref{eq:permeability}). 
The optical part tries to increase the irradiance given by the device at some given plane, typically the focal plane of the photon sieve. However, our device may present some rotational asymmetry producing  irregularities around its focal point. 
To smooth these variations, 
we define the encircle power around the focal point as the integrated irradiance on a circle on the focal plane and  centered at the focal point, $x'=y'=0$. The radius of this circle is the radius of the first minimum of the Airy spot obtained for an open circular aperture with a radius equal to the $M$th Fresnel zone, $R_{{\rm Fresnel}, M}=\sqrt{M \lambda f'}$, i.e., the radius of the aperture of the PS.
This radius is given as $r_{\rm Airy}= 0.61  \sqrt{\frac{\lambda f'}{M}}$. 
 This encircled power  is calculated as
\begin{equation}
P_{\rm PS} = \int_{S_{\rm Airy}} I_{\rm PS} (x',y') dx' dy'
\label{eq:I_airy}
,\end{equation}
where $S_{\rm Airy}$ represents the circle defined at the focal plane and having a radius $R_{\rm Airy}$.

\begin{figure}[h!]
\centering
\includegraphics[width=0.7\textwidth]{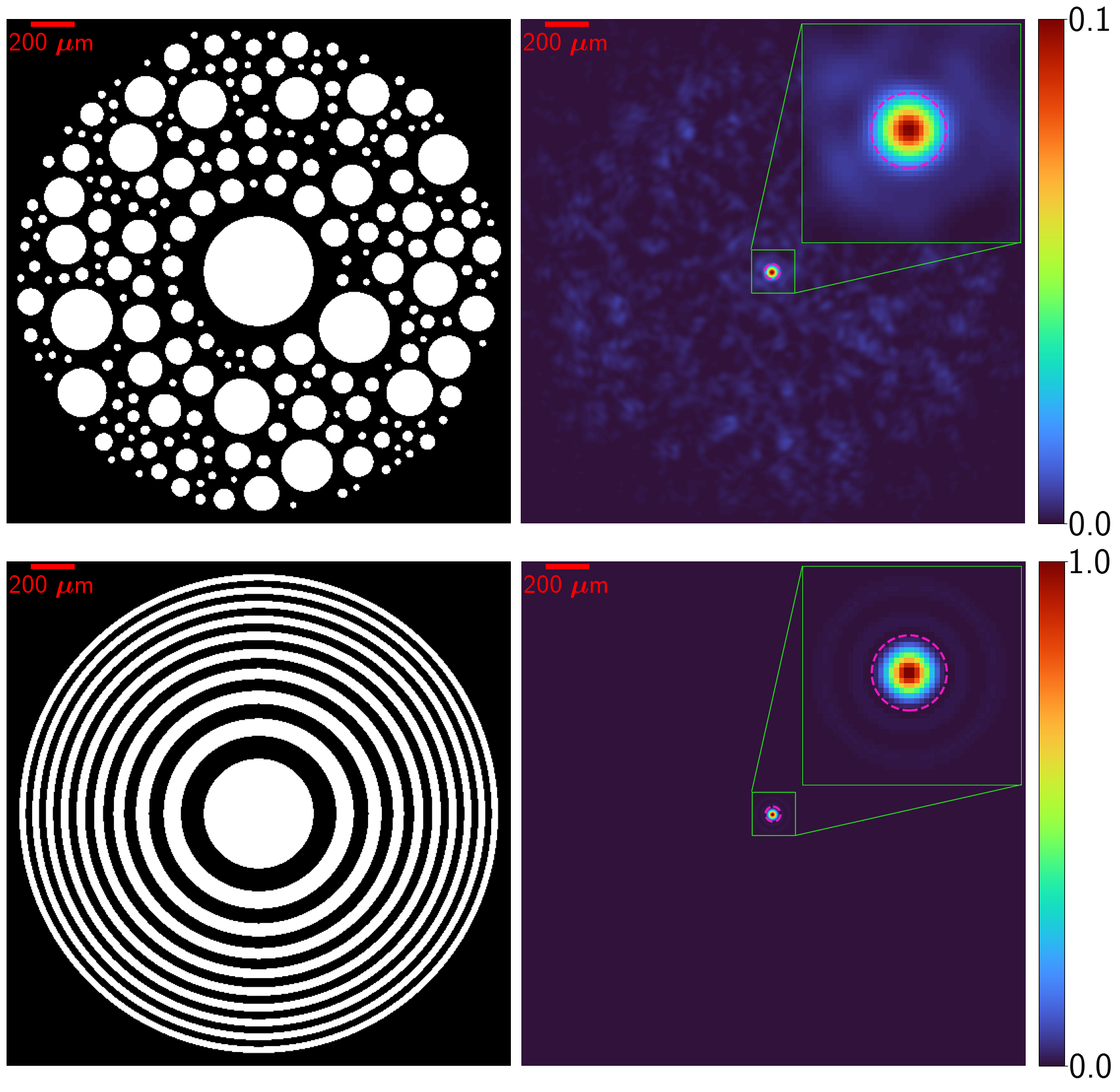}
\caption{
We can see how a PS maks (a) generates an irradiance map (b) that, for optimization reasons, is compared with the case of a Soret FZP (c) with the same focal length and aperture and opened odd zones.  
The PS mask is obtained through the HbH method presented in 
Sec. \ref{sec:hbh}.
The insets in the irradiance maps ((b) and (d)) shows the region close to the focal point. We have also added a circle to represent the region where the spatial integration to define $P^{\rm PS}$.
\label{fig:mask_and_focalplane}
}
\end{figure}

To better define the merit function from Eqs. (\ref{eq:permeability}) and (\ref{eq:I_airy}), we normalize both of them. 
The permeability of the mask, $A_{\rm PS}$,  is divided by the area of a circle having a radius, $R_{\rm PS}$, equal to $R_{{\rm Fresnel,}M}$, that is the radius of the largest Fresnel zone of the PS. This means that $A_{\rm circ} = \pi R^2_{\rm PS}$. Therefore, the normalized permeability parameter is:
\begin{equation}
\hat{A} = \frac{A_{\rm PS}}{A_{\rm circ}},
\label{eq:a_closed}
\end{equation}
where $\hat{A}$ should be maximized to increase permeability. The term of the merit function describing the optical performance of the PS, is obtained after normalizing $P_{\rm PS}$ with the encircled power obtained within the Airy disk for a perfect Fresnel lens made of open and closed zones (a Soret lens) and having $M$ zones. This ratio is given as
\begin{equation}
    \hat{P} =  \frac{P_{\rm PS}}{P_{\rm FZP}}
    \label{eq:I_normalized}
,\end{equation}
where $P_{\rm PS}$ is evaluated in Eq. (\ref{eq:I_airy}) using the irradiance pattern, $I_{\rm PS}(x',y')$, obtained from  the   PS mask. 
Figure \ref{fig:mask_and_focalplane}.a and  \ref{fig:mask_and_focalplane}.b
show the mask of a PS and the corresponding irradiance distribution, respectively. The same is represented in Fig.
\ref{fig:mask_and_focalplane}.c and \ref{fig:mask_and_focalplane}.d, for the reference Soret FZP mask.

Using these two normalized parameters defined in Eqs. (\ref{eq:a_closed}) and (\ref{eq:I_normalized}), we may define a  Weighted Merit Function (WMF):
\begin{equation}
    {\rm WMF}(\omega) = \omega \hat{P} + (1-\omega) \hat{A},
    \label{eq:weightedFoM}
\end{equation}
where $\omega$ balances the permeability and optical power terms. In our case, as we are interested in having a photon sieve with focusing capabilities, $\omega$ can not be zero because in that case the encircled power term has not influence in the design and the system would be optically useless. The optimization method looks for a maximum of  the merit function defined in Eq. (\ref{eq:weightedFoM}). 
In this contribution we  used the Nelder-Mead algorithm \cite{nelder_simplex_computerj_1965} to obtain the minimum of a function. Therefore, we add a minus sign to the definition of the merit function to properly use this algorithm.

\subsection{Geometrical Merit Function}
\label{sec:geometricalfm}

For a photon sieve made of circular apertures, the calculation of the encircled power  at the focal plane requires the evaluation of the propagated electric field and the irradiance distribution around the focal point. In the previous section we showed an analytical approach for this numerical calculation (see Eqs. (\ref{eq:amplitudemultiplecircular}) and (\ref{eq:irradiancedistribution})). However, in the following analysis we have used  the  open-source ``Diffractio" Python package \cite{Diff}, that can be easily adapted to our case and uses 
proved propagation algorithms to obtain the desired results.
In fact, ``Diffractio" implements the Rayleigh-Sommerfeld method to safely calculate the amplitude distribution after the mask.
This approach only requires the definition of the geometry and characteristics of the PS mask, $M_{\rm PS}(x,y)$, the electric field at the input plane of the mask, that we assume as an uniform plane wave,  and the wavelength, $\lambda$. 
Even though these calculations are fast, the  optimization process requires a large number of evaluations of the merit function,  with the consequence of a long time for the optimum solution to appear. This is why we also propose an alternative and faster way to generate an optimum solution. 

\begin{figure}[h!]
    \centering
    \includegraphics[width=0.8\textwidth]{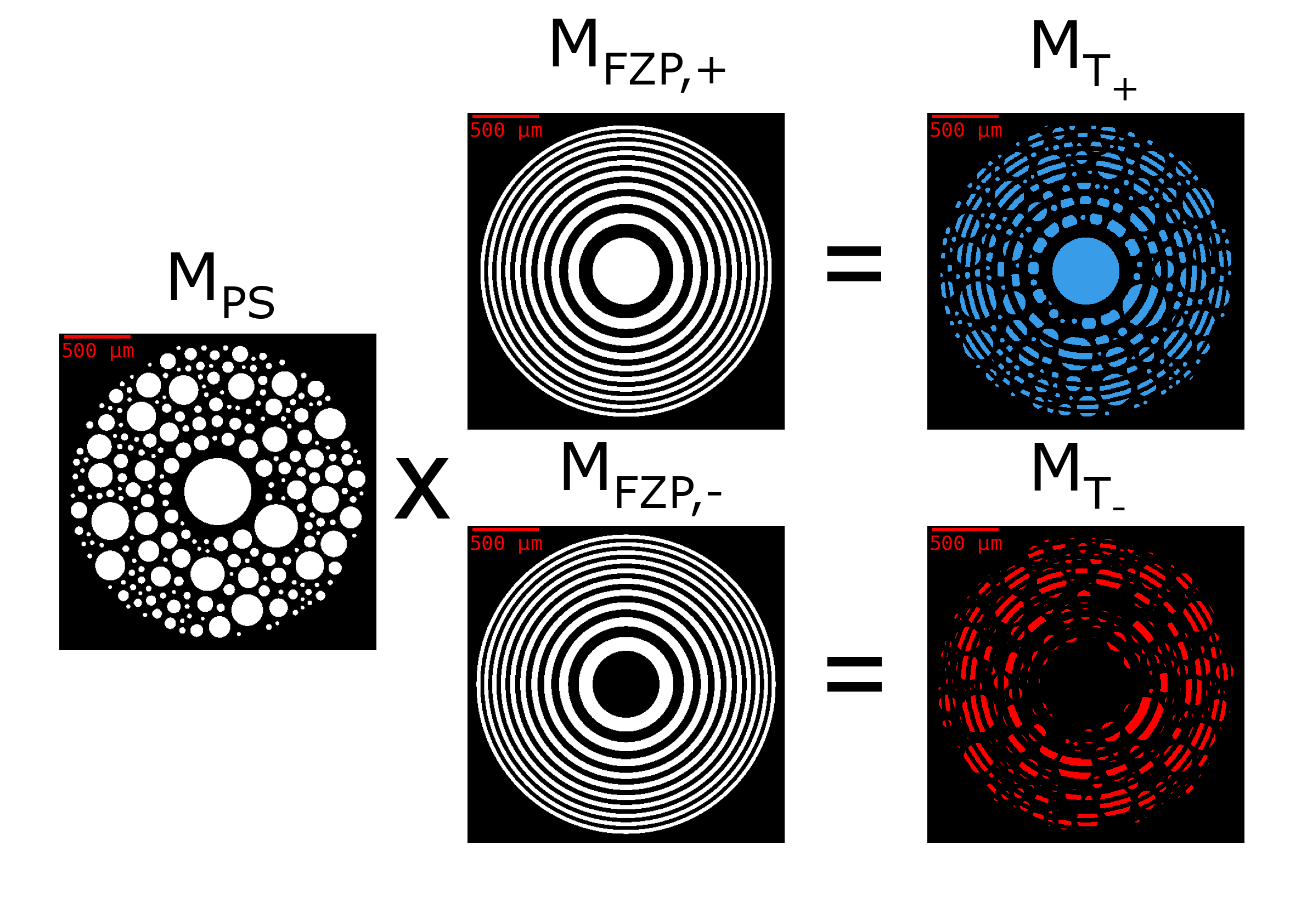}
    \caption{Graphical description for the calculation of  $T_{\pm}$ (see equations (\ref{eq:A_plus}) and (\ref{eq:A_minus})), illustrated with a photon sieve obtained using the HbH method (see section \ref{sec:hbh}.) 
$M_{{\rm S}_\pm}$ is defined as $M_{\rm PS}(x,y)  M_{{\rm FZP}_\pm}$.    
    %and $\alpha_{\pm}$ parameters
    \label{fig:DefinitionAjBj}
    }
\end{figure}

Focusing photon sieves are obtained as a variation of the well-founded FZP. In an amplitude FZP, even and odd Fresnel zones are open and closed alternatively to focus light through interference and diffraction. 
In our case, as far as we open the central, first Fresnel zone, the odd zones contribute constructively.
Every aperture of the photon sieve will intersect partially with the constructive and destructive zones.  
These contributions can be easily computed in terms of the intersection of the open area of the photon sieve and the odd and even zones, respectively for the constructive and destructive portions (see figure \ref{fig:DefinitionAjBj}) :
\begin{eqnarray}
T_{+} & = &\int_{S}  M_{\rm PS}(x,y)  M_{{\rm FZP},+} (x,y) dx dy , \label{eq:A_plus} \\
T_{-} & = & \int_{S}  M_{\rm PS} (x,y)  M_{{\rm FZP},-} (x,y) dx dy , \label{eq:A_minus}
\end{eqnarray}
where $M_{\rm PS}$ is the binary mask of the photon sieve (with ones at the open apertures and zeros in the rest), $M_{{\rm FZP,+}}$
and $M_{{\rm FZP},-} $ are the masks for the odd and even Fresnel zones, respectively,   and $S$ is the total area of the photon sieve. These two parameters can be normalized to become independent on the size of the photon sieve, as:
\begin{eqnarray}
\alpha_{+} & = & \frac{T_{+}}{T_{{\rm FZP},+}}  , \label{eq:alpha_plus}\\
\alpha_{-} & = & \frac{T_{-}}{T_{{\rm FZP},-}}  , \label{eq:alpha_minus}
\end{eqnarray}
where $T_{{\rm FZP},\pm} = \int_{S}  M_{{\rm FZP},\pm}(x,y) M_{{\rm FZP},\pm}(x,y)  dx dy $.
We combine these two dimensionless parameters into a single one that describe the net coincidence of the photon sieve mask with the constructive and destructive Fresnel zones: $\alpha = \alpha_{+} - \alpha_{-}$. This geometrical parameter allows the definition of a Geometric Merit Function (GMF) given as
\begin{equation}
{\rm GMF}= \alpha ,
\label{eq:gmf}
\end{equation}
 that does not need an evaluation of the optical irradiance at the focal plane. 

\section{Generation of Photon Sieves}
\label{sec:generation}

Once we have described the merit functions applicable to the optimization process of PSs, we describe here how they can be generated using several approaches.  
All of them begin with the central Fresnel zone open and  populate the aperture of the PS with circular apertures.
Each circular aperture is characterized by three numbers: the radius of the aperture, $R_j$, and the coordinates of its center $(x_{0,j},y_{0,j})$, where the subindex $j$ labels the aperture. The methods that we describe in this section differ in how these parameters are obtained. 
To properly compare our results, the characteristic PS parameters are the same for the four methods: $\lambda=632.8$~nm, $f'=100$~mm and an aperture radius of 
$R_{{\rm PS}}=1125$~$\mu$m (meaning that $M=20$). The fabrication limitations are parameterized as $g=28.8$~$\mu$m (see Eq. (\ref{eq:fabricationcondition})).

\subsection{Sectorized Fresnel Zones method (SFZ)}
\label{sec:fzpspoke}

When modifying an optimum amplitude FZP to become a PS, we can begin with a mask that resemble very much this design of reference. This is why, the first design in this contribution is a variation of the classical amplitude FZP that we are taking as a reference in the definitions of the merit functions. 
The idea is to sector the open zones by adding spoke (this why we call this method as Sectorized Fresnel Zones, SFZ). The main role of the spokes  is to maintain the topological continuity of the mask and avoid its collapse. We may begin with 3 spokes spaced regularly every 120$^\circ$ angularly.  The width of the spoke should be related with $g$. However, 3 spokes are not enough when moving outwards from the center of the photon sieve. To include new spokes we should establish a criteria where the new spoke begins.
As a feasible example, we can start with three spokes departing from the outer rim of the first even Fresnel zone. This radius is $R_{{\rm Fresnel},2} = \sqrt{2 \lambda f'}$. Then, the free contour between spikes can be estimated as $C_2 = \frac{2}{3} \pi \sqrt{2 \lambda f'}$. We may include the next spoke when the outer rim of the even zone is twice the outer rim of the zone where the previous spoke departs from. After applying this rule recursively, we find that the spokes duplicate, at the following  $k$-th Fresnel zones:
\begin{equation}
k = 2^{s+1} 
, \label{eq:orderspokes}
\end{equation}
where $s$ represents the successive generation of spokes along the PS. This value of $k$ determines the  radius,  $R_{{\rm Fresnel},k}$,  where the new spoke starts. 
The initial number of spokes is arbitrary and should be related with the stiffness of the material of the mask. At the same time, we have found that this initial number of spokes, that sequentically duplicates  when moving outwards, does not affect to the radial location where this duplication applies.
The mask of this sectorized Fresnel zone photon sieve is shown in 
Fig. \ref{fig:hir_sfzp}.a along with the calculated irradiance distribution at the focal plane (Fig. \ref{fig:hir_sfzp}.b)

\subsection{Hole-in-Ring Method (HiR)}
\label{sec:hir}

\begin{figure}[h!]
\centering
\includegraphics[width=0.8\textwidth]{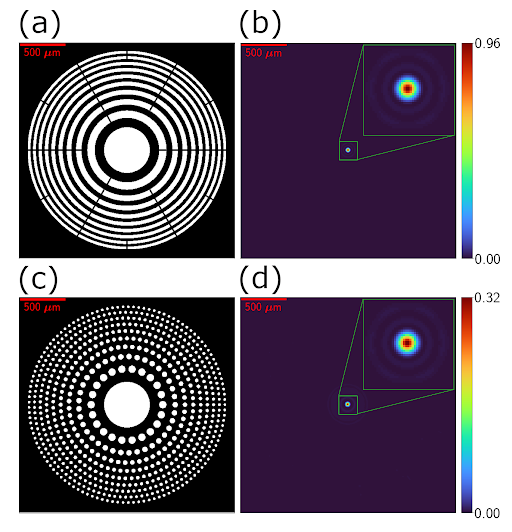}
\caption{ (a) and (b): PS mask and irradiance distribution at the focal plane obtained from the method SFZ.
(c) and (d) PS mask and irradiance distribution at the focal plane obtained from the method HiR. The color bar has a different range for the irradiance maps. The value of 1 in irradiance would correspond to the case of a perfect FZP (without spokes).
\label{fig:hir_sfzp}
}
\end{figure}

A quite simple way of filling the mask of a PS with circular apertures  is to open equal holes in every ring that contributes constructively, i.e., the odd Fresnel zones, without overlapping with even zones.
We have named this procedure as the Hole-in-Ring method (HiR).
 As it will happen with the RbR method, those holes are angularly distributed equally. The parameters of these holes are easily obtained: for the $j$-th Fresnel zone, the radii are
\begin{equation}
    R_{j} = \frac{\sqrt{\lambda f'}}{2} \left( \sqrt{j} - \sqrt{j-1}  \right)
    ,\label{eq:radius_hirmethod}
\end{equation}
and the distances of the center of these holes to the center of the PS are
\begin{equation}
    r_j = \sqrt{ x_{0,j}^2 + y_{0,j}^2} = \frac{\sqrt{\lambda f'}}{2} \left( \sqrt{j} + \sqrt{j-1}  \right)
    .\label{eq:center_hirmethod}
\end{equation}
A realization of this photon sieve can be seen in Fig. \ref{fig:hir_sfzp}.c. Along the positive $x$ direction we can see how the holes have their centers aligned. This situation can be compared with the mask presented in Fig. \ref{fig:RbR_photonsieve}.b where the angular location is shifted from ring to ring to avoid this effect. The effect on the irradiance distribution is not relevant. This irradiance distribution can be seen in Fig. \ref{fig:hir_sfzp}.d.  When comparing the irradiance maps in this figure, we can notice how the color bar has a different range, meaning that the SFZ design generate more irradiance than the HiR method.

\subsection{Iterative Ring-by-Ring Method (RbR)}
\label{sec:rbr}

In this subsection we propose the Ring-by-Ring method (RbR) which  generates always the same solution, except for an angular location shift that will be discussed later.
The method is based on an iterative algorithm that, from the point of view of the parameters of interest, works along a radius of the PS mask. 
The first aperture is also the central one, that could be interpreted as the first ring having only one aperture. Every ring will have equal circular apertures regularly spaced  in angle. 
The key point of the algorithm is to  choose the appropriate value of the radius of these equal apertures and the radial position of its center. To do that, we generate a map of the merit function in terms of these two variables.  On the other hand,
the merit function requires the calculation of the irradiance at the focal region, and the total opened aperture. To do that, we fill one ring at the time  with a collection of equal apertures angularly spaced. 
This evaluation also considers the limitation given by equation (\ref{eq:fabricationcondition}) for the radial and angular coordinates. 
After choosing those parameters providing the optimum performance, we obtain the parameters of the circular apertures that  are arranged as a ring. 
This procedure is repeated, including the evaluation of the merit function, until filling the aperture of the PS. 
 
\begin{figure}[h!]
\centering
\includegraphics[width=0.8\textwidth]{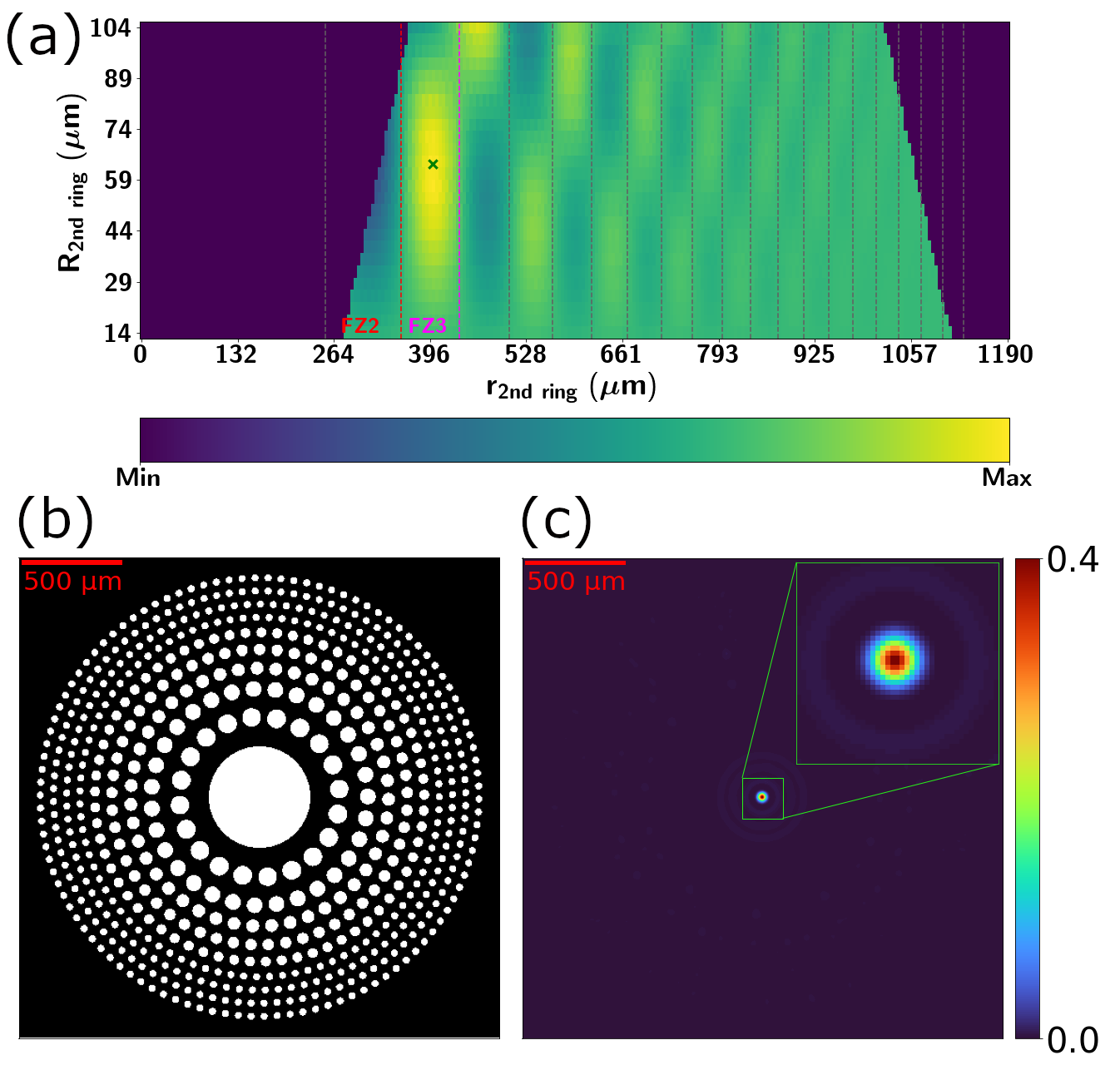}
\caption{
(a) Map of the merit function in terms of the radius of the aperture, $R_{{\rm 2}^{\rm nd} {\rm ring}}$, and its radial location, $r_{{\rm 2}^{\rm nd} {\rm ring}}$, for the case of having only the first central aperture. This evaluation serves to calculate the parameters of the second ring. The range in the colorbar expands from the minimum to the maximum values of the merit function for those combination of radial position and hole radius available after applying the fabrication constrains.  The vertical lines represent the limits of the Fresnel zones, $R_{{\rm Fresnel},j}$. The value of the maximum of the merit function is marked with a cross. 
(b) Mask of the photon sieve obtained for $\lambda=632.8$~nm, $f'=100$~mm, and aperture size $R_{{\rm PS}}\approx 1125~\mu$m  ($M=20$). 
(c) Spatial distribution of the irradiance at the focal plane of the PS.
\label{fig:RbR_photonsieve}
}
\end{figure}

In Fig. \ref{fig:RbR_photonsieve}.a we illustrate how this method works when determining the parameters of the apertures of the second ring (only the central aperture is open) as a function of the radial location of the center of the apertures, $r_{{\rm 2}^{\rm nd} {\rm ring}}$, and their radius, 
$R_{{\rm 2}^{\rm nd} {\rm ring}}$. The map shows the WMF with a value $\omega=1$, meaning that we are only interested in maximizing the irradiance.  
The trapezoidal shape of the region where WMF is evaluated describes the effect of the fabrication constrains. It means that larger circles have to be placed farther from the central aperture and from the outer rim of the PS. We have also represent the Fresnel zones with vertical dotted lines. The maximum of the merit function appears at the third zone and is marked with a cross.  Each iteration to obtain a new ring, requires the evaluation of the merit function within an availability region determined by the parameters of the previous iterations and the fabrication limitations. 
 The final result of this method is presented as a mask in Fig. \ref{fig:RbR_photonsieve}.b. The corresponding map of the irradiance at the focal plane of the PS is given in Fig. \ref{fig:RbR_photonsieve}.c.

For a preset merit function, the RbR method always provides the same  optimum solution which is 
determined by the design parameters: wavelength, focal length, aperture, and fabrication constrains ($g$ and $R_{\rm min}$).
In this  method there exist a free parameter that may generate slightly different photon sieves. This parameter is the angular location of the apertures along different rings. As far as the optimization problem only considers the distance between the new aperture and the center, this center can be angularly aligned along the same radius of the photon sieve, or change from ring to ring, as we did in our designs. In any case, the effect of this angular arrangement in the irradiance at the focal plane is negligible. The effect of this angular shift can be seen when comparing the alignment of the circular apertures along the positive $X$ axis of the masks in Figs. \ref{fig:RbR_photonsieve}.b, and \ref{fig:hir_sfzp}.a as we  briefly discussed in subsection \ref{sec:hir}.

\subsection{Iterative Hole-by-Hole Method (HbH)}
\label{sec:hbh}

This method, named as Hole-by-Hole method (HbH), iteratively opens a single hole, one after the other, until filling the aperture of the photon sieve, or some other stop conditions related with the allocated computational time  are met. 
Sparse and randomly distributed apertures are often used to improve quality and resolutions in remote sensing when a collection of small aperture telescopes are used together
\cite{fiete_sparseaperture_aiprw2000,introne_spareaperture_phthesis2004}.
The HbH approach is based on the limitations of the available area to place the $j^{\rm th}$ hole after complying with the fabrication condition, Eq. (\ref{eq:fabricationcondition}), for all the previously opened apertures ($i=1,\ldots,j-1$). The method is iterative, and in each iteration places a hole on the PS mask. 
It also uses a hierarchy when choosing the order of the parameters of the hole. 
First, we select the value of the radius of the new hole, $R_j$, from a range of values that lies in a given range $[R_{j,{\rm min}}, R_{j,{\rm max}}]$. 
$R_{j,{\rm min}}$ corresponds with the radius of the smaller fabricable hole (Eq. (\ref{eq:rho_min})). 
However, although $R_{j,{\rm max}}$
departs from the value given in Eq. (\ref{eq:rho_max}), it
  decreases with the number of iterations, meaning that larger apertures are harder to place as the photon sieve is populated with more and more holes. 
The first choice for $R_j$ is a random value between $R_{j,{\rm min}}$  and $R_{j,{\rm max}}$. 
Once the radius of the hole is set, the algorithm maps the aperture of the photon sieve to know the positions fulfilling condition (\ref{eq:fabricationcondition}), and randomly choose one of the available locations. Then, the merit function is computed and compared with its previous evaluation looking for an optimization. The search for the aperture $j$ finishes when the time computation exceeds an amount previously defined, or the number of iterations reaches the number of holes previously set.  If the computation time and the desired number of holes are below their preset values but there is no space in the mask, then the algorithm does not put any more holes and stops.

\begin{figure}[h!]
    \centering
    \includegraphics[width=0.8\textwidth]{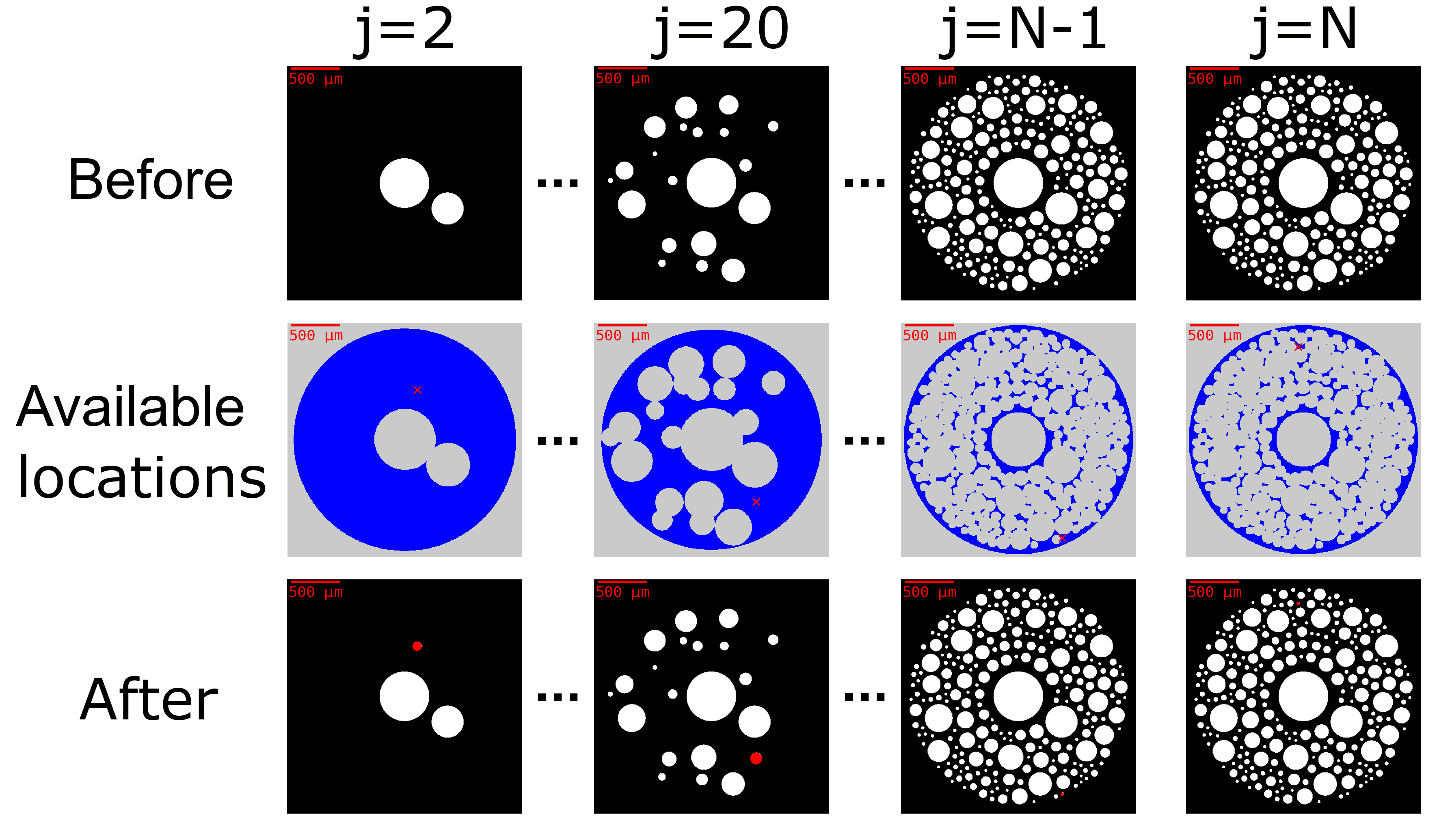}
    \caption{
    Evolution of the hole-by-hole method for several values of $j$ ($j=2,20,N-1,N$). The first row depicts the photon sieve mask for the iteration $j-1$, the second row represents the available area of the photon sieve for the radius $R_j$ and the position for the next hole (red cross). The third row corresponds with the result of the iteration $j$ after the hole is placed (we have marked the new hole in red).
    \label{fig:HbH_photonsieve}
    }
\end{figure}

The result is a
collection of $N$ holes located on the PS mask. 
As far as the method is governed by a random choice, each realization of this
algorithm will produce a different mask. 
It is interesting to note that an apparent random collection of holes is able to focus light at the desired location, after an educated choice of the location and size of the holes. 
In figure \ref{fig:HbH_photonsieve} we show four iterations of one of the realizations of the algorithm. The first row shows the situation before the $j$ step, the second row shows the map of the available locations for the aperture having a radius $R_j$, and the third row correspond with the photon sieve mask after placing this aperture at the location generated by the algorithm. 

\section{Analysis and Results}
\label{sec:analysis}

The design strategies shown in Sec. \ref{sec:generation} generate a collection of photon sieves that perform differently. 
In section \ref{sec:optimization} we  defined a way of evaluating the merit function from the aperture distribution,  $M(x,y)$. From the point of view of the optimization, only those masks generated through methods HbH and RbR are iterative and can be studied in terms of the iterations that generate additional apertures to the PS.
 Actually, as previously explained in section \ref{sec:rbr}, once we set the merit function, RbR method always produces the same solution, leaving the HbH method as the one deserving a detailed analysis of the different realizations.
Due to the additive generation approach of HbH and RbR methods, the permeability term always grow with the number of iterations.
On the contrary, the encircled power term in HbH may decrease when forcing the algorithm to place a new aperture. Then, HbH method is open to multiple variations when considering how to stop the algorithm at an acceptable performance. 
Table \ref{tab:irradiance_and_permeability} summarizes the values of the ratio of the irradiance, $\hat{P} = P_{\rm PS}/P_{\rm Fresnel}$, 
and open areas, $\hat{A} = A_{\rm PS}/A_{\rm circ}$ (see Eqs (\ref{eq:a_closed}) and 
(\ref{eq:I_normalized})) where the best performing mask having circular apertures is hihglighted in gray. 
We have included an uncertainty value for the HbH method only, because it is the one that generates different masks for every realization. 
Table \ref{tab:irradiance_and_permeability} also shows the average time of computation per iteration for the HbH method. We can see that the GMF is about 3 times faster than the WMF method.

\begin{table}[h!]
    \caption{Comparative values of the encircled power, $\hat{P}$, permeability, $\hat{A}$, and computation speed (as the number of second for iteration) for the methods analyzed in this contribution.     
    \label{tab:irradiance_and_permeability}
}
    \centering
    \begin{tabular}{rccc}
    %\hline 
      Method   &  $\hat{P}$ &  $\hat{A}$ & Speed [s/iter]\\
      \hline 
	  SFZ   				  &    0.961       		  & 	   0.44       & --      \\      
	  HiR   				  &    0.379        	  &        0.23       & --      \\
      \rowcolor[HTML]{E5EBEF}
      RbR   				  &    0.558         	  &        0.36 	  & --      \\     
      HbH (WMF, $\omega = 1.0$) & $0.117  \pm 0.007$    & $0.44 \pm 0.01$     & $25 \pm 2$ \\  
      HbH (WMF, $\omega = 0.7$) & $0.090  \pm  0.007$    & $0.52 \pm 0.02$     & $34 \pm 4$\\
      HbH (GMF)   			              & $0.14    \pm  0.01  $    & $0.42  \pm  0.02$   & $9  \pm 3$ \\
      \hline
    \end{tabular}
\end{table}

\begin{figure}[h!]
\centering
\includegraphics[width=0.95\textwidth]{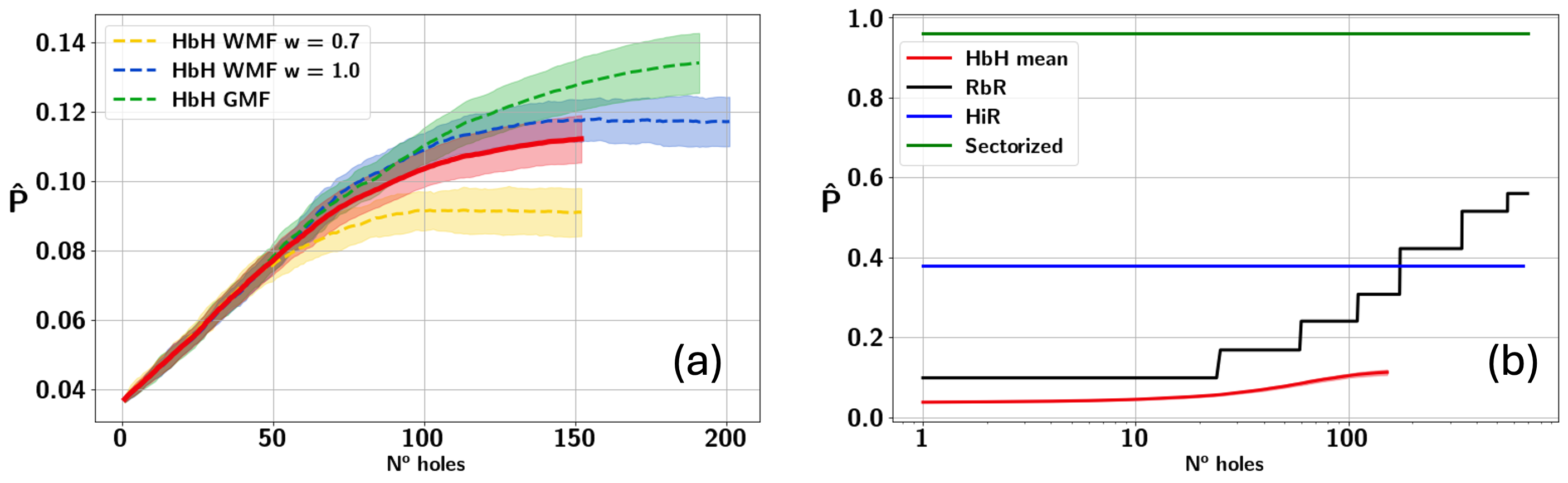}
\caption{
(a) Comparison of four collections of 73 PS realizations generated by the HbH method using three merit functions: WMF with $\omega=0.7$ (in yellow) and $\omega=1.0$ (in blue), and  GMF (in green). The solid red line corresponds to the average of these three cases. 
(b) Evolution in terms of the iteration number of the ratio $\hat{P}=P_{\rm Airy}/P_{\rm Fresnel}$ defined in Eq. (\ref{eq:I_normalized}), for photon sieves obtained by the HbH method. 
The red line represents the average of these realizations. The black line shows the evolution of the same ratio for the RbR method, the stair-like shape of this plot happens because all the apertures in the same ring are obtained at the same time. The blue and green horizontal lines represent the cases of the HiR and SFZ methods, respectively. The number of hole axis is represented in a log scale to better show the four methods. 
\label{fig:iterative_analysis}
}
\end{figure}

Figure 
 \ref{fig:iterative_analysis}.a shows the value of the normalized encircled energy parameter, $\hat{P}$, as a function of the iteration steps for the method HbH. 
 We may see
 how this parameter strongly increases during the first iterations  before reaching a stable value. 
 The realizations of the method produce different masks with variations of the values of the irradiance, that is represented as shaded regions with a width equal to the standard deviation.  As expected, when the aperture factor is considered ($\omega=0.7$), the value of the irradiance diminishes (see the yellow line and region) with respect to the case when 
 $\omega=1.0$, represented in blue. 
At the same time, we see that the geometrical merit function in the HbH method generates PS with larger values of the irradiance and permeability factors.  

Fig. \ref{fig:iterative_analysis}.b summarizes the performance of the PS in terms of the value of $\hat{P}$ for the four methods. 
The horizontal lines represent the case of the SFZ (in green) and the HiR (in blue). As far as all their open apertures are obtained simultaneously, we have represented them as a constant value, being the SFZ case the one behaving the best because it is the design closest to the reference FZP. 
The RbR method add apertures when each additional ring is analyzed. Then, the dependence of  $\hat{P}$ vs. number of holes, has a staircase shape with steps corresponding to the added ring. Finally, the HbH methods provides PS with lower encircled energy. We have represented in red the average of this parameters also plotted in Fig, \ref{fig:iterative_analysis}.a.
  Overall,  the irradiance values  for the HbH method are lower than those obtained by other methods (as shown in table \ref{tab:irradiance_and_permeability}).

\begin{figure}[h!]
    \centering
    \includegraphics[width=0.8\textwidth]{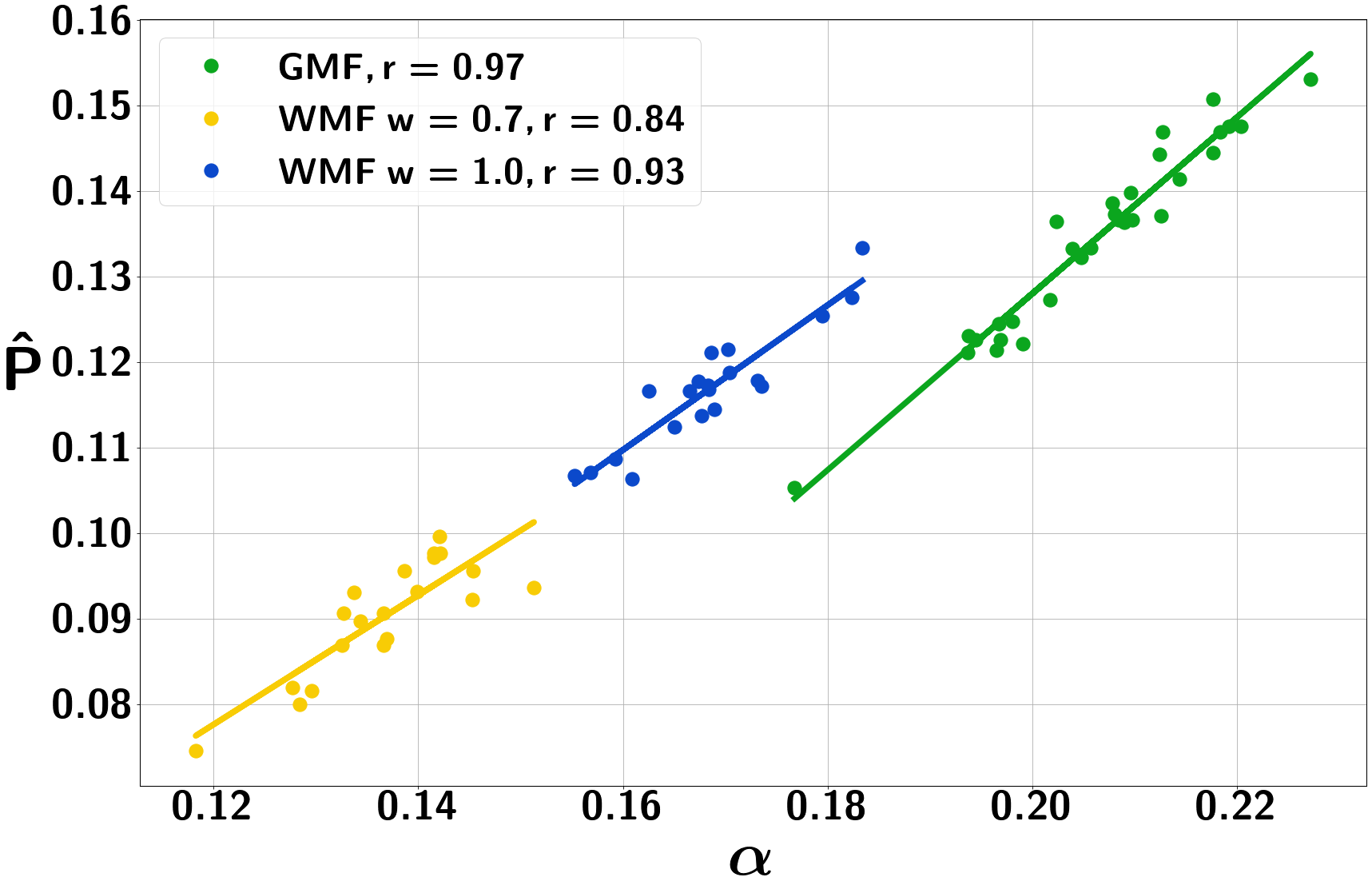}
    \caption{Evolution of the encircled power ratio, $\hat{P}$ in terms of the values of $\alpha$. We have plotted in this graph three cases of PS obtained using the HbH method. These cases  differ in the definition of the merit function. The red dots and line correspond with a merit function that maximizes the irradiance at the Airy disk region. The blue dots and line are for a combination of both the irradiance and the permeability factor (see Eq. (\ref{eq:weightedFoM})) when $w=0.7$. Finally, the green dots and line correspond with the geometric merit function (see Eq. (\ref{eq:gmf})).
        \label{fig:irradiance_vs_alpha}
}
\end{figure}

From its definition in Eq. (\ref{eq:gmf}), the geometric merit function is also described by $\alpha$. 
In Fig. \ref{fig:irradiance_vs_alpha} we show the irradiance parameter as a function of $\alpha$ (see the definition of $\alpha$ in terms of $\alpha_-$ and $\alpha_+$ in Eqs. (\ref{eq:alpha_minus}) and (\ref{eq:alpha_plus})). 
The three examples of the merit function are represented in different color. The case of WMF with $\omega=0.7$ (in yellow) is the one generating a lower value of $\hat{P}$. When $\omega=1.0$ (in blue), the encircled energy and the resemblance with the ideal FZP arrangement grows. This is because the merit function balance the maximization of $\hat{P}$ and $\hat{A}$ in a quite different way: if $\omega=1$, the optimization forgets about $\hat{A}$, meanwhile, for $\omega=1$, only $\hat{P}$ is maximized.
We may see that both $\alpha$ and 
$\hat{P}$  are larger for the case of the GMF (in green), meaning that this approach generates masks with closer resemblance to the FZP,  showing a larger value of the encircled power. 
This plot also shows that both parameters, $\alpha$ and $\hat{P}$, are 
proportional with linear correlation factors reaching a value of $r= 0.97$ for the case of the PS generated using the GMF.

\section{Conclusions}

\label{sec:conclusions}

In this contribution we present several reliable methods to generate focusing and permeable photon sieves. 
All these methods are constrained by fabrication limitations regarding the maximum number of usable Fresnel zones, and the minimum value of the radius of the involved apertures. 
Using an amplitude Fresnel lens as reference, we have obtained useful relations between these limitations an the optical characteristics of the fabricable photon sieves.
Our design methods generate photon sieves that can be optimized balancing the focusing capabilities of the element, given in terms of the encircled power located around the focal point at the focal plane, and the permeability of the
mask, given in terms of its open area.
As far as the optimization of the design is key for our designs, we have defined and implemented a weighted function of merit that can be calculated by using a propagation algorithm, or by evaluating an analytical solution able to produce the irradiance distribution at the focal plane. 
A geometrical version of the merit function which does not require to propagate the beam is also defined in terms of the
similarities of the PS mask with the odd and even Fresnel zones of a FZP. The advantage of this
geometrica merit function is that it does not require to propagate the beam from the aperture to the focal plane, saving  time when evaluating the merit function.  The results show a better optical performance, being this evaluation 3 times faster than those based on the weighted merit function.

Departing from the classical FZP as a reference, we have designed a sectorized FZP arrangement by including spokes regularly spaced. These spokes help to maintain the physical integrity of the
mask. When moving to circular apertures photon sieves, we have obtained a simple design, named as 
Hole-in-Ring, where the circular apertures are opened in each even Fresnel zone locating their centers in the middle of the zone and with a diameter equal to the width of the given zone. These two previous design do not require any optimization can be considered as one of the simplest variations for a focusing photon sieve. 
However, this contribution explores the optimization design using previously defined merit functions. Then, once the merit function is set, the Ring-by-Ring method is a deterministic approach that evaluates the merit function for every possible choice of the involved parameters, assuming an approach where the parameters of the apertures populating each ring are determined iteratively. However, in this design, the apertures at each ring are equal and  regularly spaced across the
$2\pi$ angular domain. Finally, the Hole-by-Hole method is a quite different and iterative method that opens an aperture at the time and relies on a random, but guided, choice of the characteristics parameters of the new aperture. This method generates a different
mask everytime it runs. Besides the intrinsic randomness of this method, the possible variations in the output are also related to the choice in $\omega$ for the weighted merit function or the geometrical merit function. 

To compare the solutions given by these approaches, we have calculated the encircle power and permeability parameters for several cases of interest. 
For the HbH method we have obtained a relation between the optical performance portion of the merit function and the 
value of a simple geometrical parameters that can be easily calculated. 
Our results show that a photon sieve generated with circular apertures, and designed using the RbR method, performs the best for the cases treated here.

In summary, using the methods provided in this paper, it is possible to generate a valid mask that works efficiently as a permeable diffactive optical element. This type of elements can be used to include optical elements within running fluids, as a part of a monitoring or sensing system. 

\section*{Acknowledgments}

This work has been funded by the projects "Nanorooms" and "VDOEST" with references PID2019-105918GB-I00 and PID2022-138071OB-I00, respectively, from the Ministerio de Ciencia e Innovación of Spain.

\bibliographystyle{elsarticle-num} 
\bibliography{librarypaper}

\end{document}